\documentclass[3p,twocolumn]{elsarticle}

\usepackage{amsfonts,amsmath,amssymb}
\usepackage{graphicx,color}

\def\blfootnote{\xdef\@thefnmark{}\@footnotetext}

\newcommand{\gsim}{\gtrsim}

\begin{document}

\begin{frontmatter}

\title{Heavy Flavor at the Large Hadron Collider in a Strong
        Coupling Approach}

\author[a]{Min He}
\author[a]{Rainer J.\ Fries}
\author[a]{Ralf Rapp}

\address[a]{Cyclotron Institute and Department of Physics \& Astronomy,
Texas A\&M University, College Station, TX 77843, USA}

\date{\today}

\begin{abstract}
Employing nonperturbative transport coefficients for heavy-flavor (HF)
diffusion through quark-gluon plasma (QGP), hadronization and hadronic matter,
we compute $D$- and $B$-meson observables in Pb+Pb ($\sqrt{s}$=2.76\,TeV)
collisions at the LHC. Elastic heavy-quark scattering in the QGP is
evaluated within a thermodynamic $T$-matrix approach, generating resonances
close to the critical temperature which are utilized for recombination into
$D$ and $B$ mesons, followed by hadronic diffusion using effective hadronic
scattering amplitudes. The transport coefficients are implemented via
Fokker-Planck Langevin dynamics within hydrodynamic simulations of the
bulk medium in nuclear collisions. The hydro expansion is quantitatively
constrained by transverse-momentum spectra and elliptic flow of light
hadrons. Our approach thus incorporates the paradigm of a strongly coupled
medium in both bulk and HF dynamics throughout the thermal evolution of the
system.

\end{abstract}


\end{frontmatter}

\section{Introduction}
\label{sec_intro}
Open heavy-flavor (HF) observables have developed into a key probe of the
hot nuclear medium produced in ultrarelativistic heavy-ion collisions
(URHICs)~\cite{Rapp:2009my}. Once charm ($c$) and bottom ($b$) quarks are
produced in primordial nucleon-nucleon collisions, their large masses
suppress inelastic re-interactions, rendering their subsequent diffusion a
quantitative tool to determine the thermalization timescale in the medium.
Since this timescale appears to be comparable to the typical lifetime of
the fireball formed in URHICs, the modifications imprinted on the
final HF spectra provide a direct measure of the coupling strength to the
medium.

The discovery~\cite{Abelev:2006db,Adare:2006nq,Adare:2010de} of the
suppression of HF decay electrons in Au+Au collisions
at the Relativistic Heavy Ion Collider (RHIC), accompanied by a remarkable
elliptic flow, has become a benchmark of our understanding of the
strongly coupled quark-gluon plasma (sQGP).
The results imply substantial thermalization of heavy quarks due to
frequent rescattering on medium constituents~\cite{vanHees:2005wb}, with
an estimated diffusion coefficient of ${\cal D}_s\simeq 4/(2\pi T)$.
Several transport models have been developed to scrutinize these
findings~\cite{vanHees:2005wb,Moore:2004tg,Zhang:2005ni,vanHees:2007me,Gossiaux:2009mk,Akamatsu:2008ge,Mazumder:2011nj,Uphoff:2011ad,Alberico:2011zy,He:2011qa,He:2012df},
differing in the microscopic interactions (both perturbative and
non-perturbative), the modeling of the bulk medium evolution (fireballs,
hydrodynamics and transport simulations), and the treatment of the HF
kinetics (Boltzmann or Langevin simulations). With the advent of Pb+Pb
collisions at the Large Hadron Collider (LHC), HF probes have entered
a new era. The more abundant production of heavy quarks makes direct
information on HF mesons available, allowing to disentangle
charm and bottom spectra. The ALICE data corroborate a strong
suppression and large elliptic flow of $D$ mesons and non-photonic
electrons in Pb+Pb($\sqrt{s_{\rm NN}}$=2.76\,TeV)
collisions~\cite{ALICE:2012ab,Abelev:2013lca,Sakai:2013ata}, but their simultaneous
description is not easily achieved by existing theoretical
models~\cite{Aichelin:2012ww,Uphoff:2012gb,He:2012xz,Lang:2012cx,Alberico:2013bza,Cao:2013ita,Armesto:2005iq,He:2011pd,Buzzatti:2011vt}.
Non-prompt $J/\psi$, associated with $B$-meson decays, measured by
CMS~\cite{Chatrchyan:2012np,CMS:2012wba} have opened a window on
bottom-quark interactions with the medium. Finally, ALICE has presented
first data on $D_s$ mesons in Pb+Pb~\cite{Innocenti:2012ds}, which have been
suggested as a particularly valuable probe to disentangle QGP and hadronic
effects in the HF sector~\cite{He:2012df}.

In the present paper we conduct a systematic comparison of our earlier
constructed transport approach for open HF~\cite{He:2011qa,He:2012df} to available
observables at the LHC. This approach implements a strong-coupling
scheme in both micro- and macro-physics (i.e., HF transport and bulk
evolution, respectively) of QGP and hadronic matter, and has been
found to describe HF data at RHIC fairly well~\cite{He:2011qa,He:2012df}.
Its building blocks are a quantitatively
constrained hydrodynamic bulk evolution~\cite{He:2011zx} into which HF
transport is implemented using nonperturbative interactions for heavy
quarks~\cite{Riek:2010fk} and mesons~\cite{He:2011yi} through QGP,
hadronization~\cite{Ravagli:2007xx} and hadronic phases of a nuclear
collision.  Since the diffusion processes are restricted to elastic
interactions, it is of particular interest to study whether the much
increased $p_T$-reach at the LHC requires additional physics not included
in our approach, e.g., radiative processes. The predictive power of our
calculations is retained by utilizing microscopic HF transport coefficients without
$K$-factors. For the application to LHC we have refined our earlier
reported results~\cite{He:2012xz} with improved heavy-quark (HQ) baseline
spectra and fragmentation in $pp$ collisions, an update of the HQ $T$-matrix
by including the gluonic sector, and a revised tune of the hydrodynamic model
to bulk observables.

In the following, we first briefly review our nonperturbative diffusion
framework emphasizing the updated inputs (Sec.~\ref{sec_form}). We then
present comprehensive HF results for $D$, $D_s$, $B$ mesons and decay
electrons in Pb+Pb (2.76\,TeV), and compare them to available
data (Sec.~\ref{sec_obs}). We summarize in Sec.~\ref{sec_sum}.

\section{Non-Perturbative HF Transport}
\label{sec_form}
Our formalism for HF transport through QGP, hadronization
and hadronic phase has been introduced in Ref.~\cite{He:2011qa}.
We here recollect its main components and elaborate on updated
inputs adequate for the phenomenology at LHC.

We compute the space-time evolution of the heavy-quark (-meson) phase-space
distribution in the QGP (hadronic matter) using the Fokker-Planck (FP)
equation, implemented via Langevin dynamics~\cite{He:2013zua}.
The FP equation follows from the Boltzmann equation through a second-order
expansion in the momentum transfer, $k$, which is justified for HF momenta
satisfying $p^2 \sim m_Q T \gg T^2 \sim k^2$ ($m_Q$: HQ mass); the pertinent
Einstein equation has been verified for nonperturbative interactions in
Ref.~\cite{vanHees:2004gq}. The FP equation encodes
the diffusion properties in well-defined transport coefficients which
can be computed from in-medium scattering amplitudes without the notion of a
cross section~\cite{He:2013zua}.

In the QGP, the thermal relaxation rates, $A(p,T)$, for heavy quarks
are taken from a thermodynamic $T$-matrix approach~\cite{Riek:2010fk},
which utilizes input potentials from thermal lattice QCD (lQCD), properly
corrected for relativistic effects and consistent with HF spectroscopy
in vacuum. In the QGP, we focus on the results using the internal energy,
as the pertinent $T$-matrices lead to better agreement with the
(independent) thermal lQCD ``data" for quarkonium correlators and HQ
susceptibilities~\cite{Riek:2010fk,Riek:2010py}.
In our previous studies~\cite{He:2011qa,He:2012df,He:2012xz},
nonperturbative HQ scattering off light quarks was supplemented with
perturbative scattering off gluons; here, we replace the latter by the
recently calculated HQ-gluon $T$-matrices~\cite{Huggins:2012dj}, with the
same lQCD potentials as for heavy-light quark interactions. This improvement
leads to a ca.~25\% increase of the total HQ relaxation rate. In addition, we
allow for a further increase of $\sim$20\% to represent the uncertainty when
going from the HQ internal energies of Ref.~\cite{Kaczmarek:2005ui} (used
in our previous calculations) to those of Ref.~\cite{Petreczky:2004pz},
cf.~Ref.~\cite{Riek:2010fk}.
The resulting HQ relaxation rates are enhanced over leading-order
perturbative calculations~\cite{Svetitsky:1987gq} (with $\alpha_s$=0.4)
by up to a factor of $\sim$5 at low momenta and temperatures close to
$T_{\rm c}$. This enhancement is caused by near-threshold resonance
structures which develop close to $T_{\rm c}$; it is reduced at higher
$T$ (e.g., to a factor of 2.5-3 at 2\,$T_{\rm c}$) and at higher momenta
where the perturbative results are
approached. This dynamical 3-momentum dependence will be relevant in the
$D$-meson nuclear modification factor discussed below. The resulting
HQ spatial diffusion coefficient ${\cal D}_s=T/(m_QA(p=0,T))$ turns out to
be quite comparable to quenched lQCD data~\cite{Ding:2011hr,Banerjee:2011ra}.

Around a pseudo-critical temperature of $T_{\rm pc}$=170\,MeV, heavy
quarks are hadronized into HF hadrons using the resonance recombination
model (RRM)~\cite{Ravagli:2007xx}, with $p_t^Q$-dependent rates taken
from the in-medium $T$-matrix. For simplicity, we here neglect the effects of a finite recombination time
window~\cite{He:2012df} which could reduce the final $D$-meson $v_2$ by up to 10\%.
The ``left-over" quarks are fragmented (see below).
In the RRM part of the $D$ and $B$ spectra, we account for the
difference of hadro-chemistry in AA and $pp$ collisions as in
Ref.~\cite{He:2012df}; most notably, the strangeness enhancement in
AA enhances $D_s$ and $B_s$ prodction, thereby slightly reducing the
fraction of charm in $D$ and bottom in $B$ mesons.

In hadronic matter, the diffusion of $D$ and $B$ mesons is continued
with transport coefficients calculated from elastic scattering amplitudes
off pions, kaons, etas, anti-/nucleons and anti-/deltas~\cite{He:2011yi}.

\begin{figure} [!t]
\includegraphics[width=1.05\columnwidth]{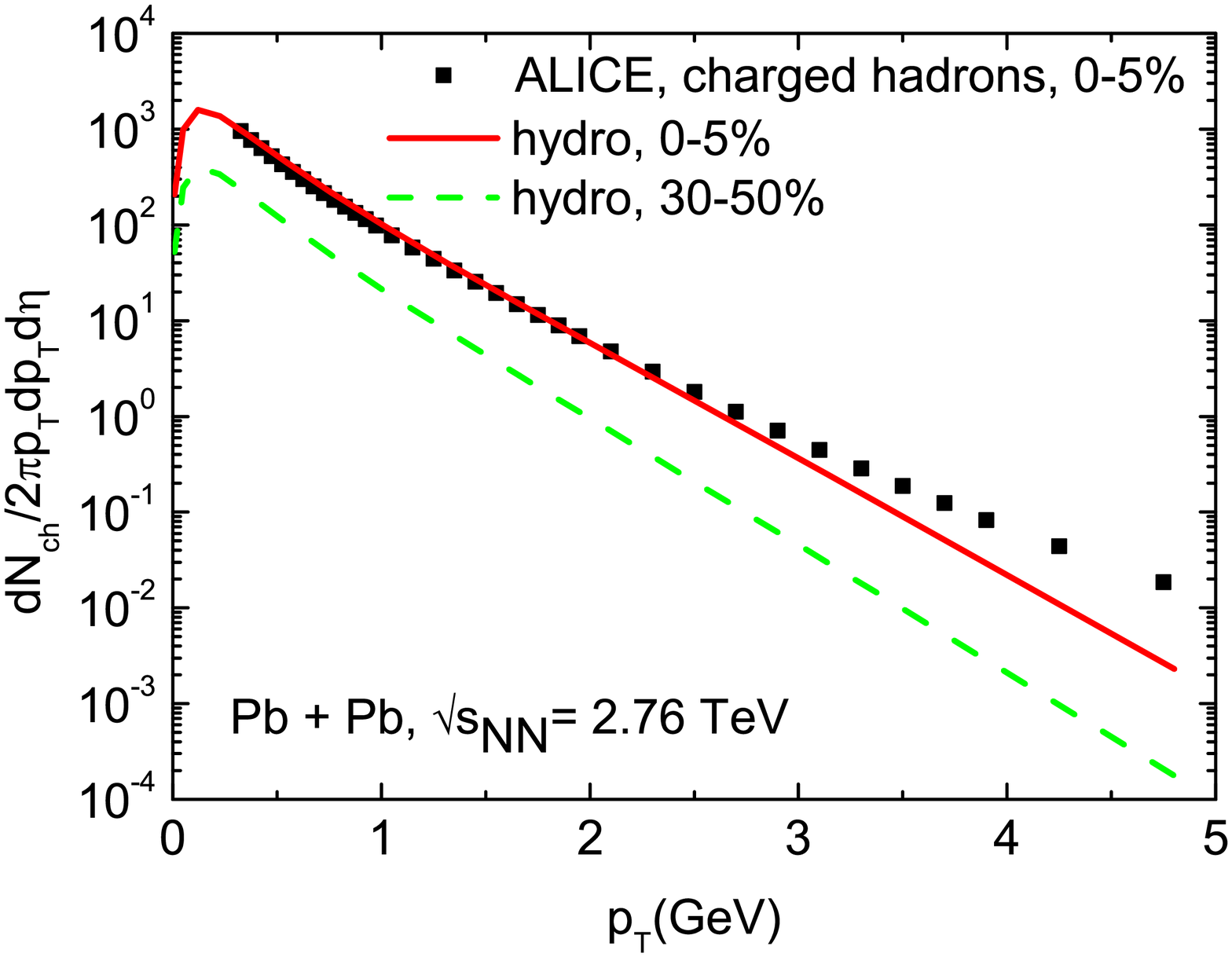}

\vspace{-0.5cm}

\includegraphics[width=1.05\columnwidth]{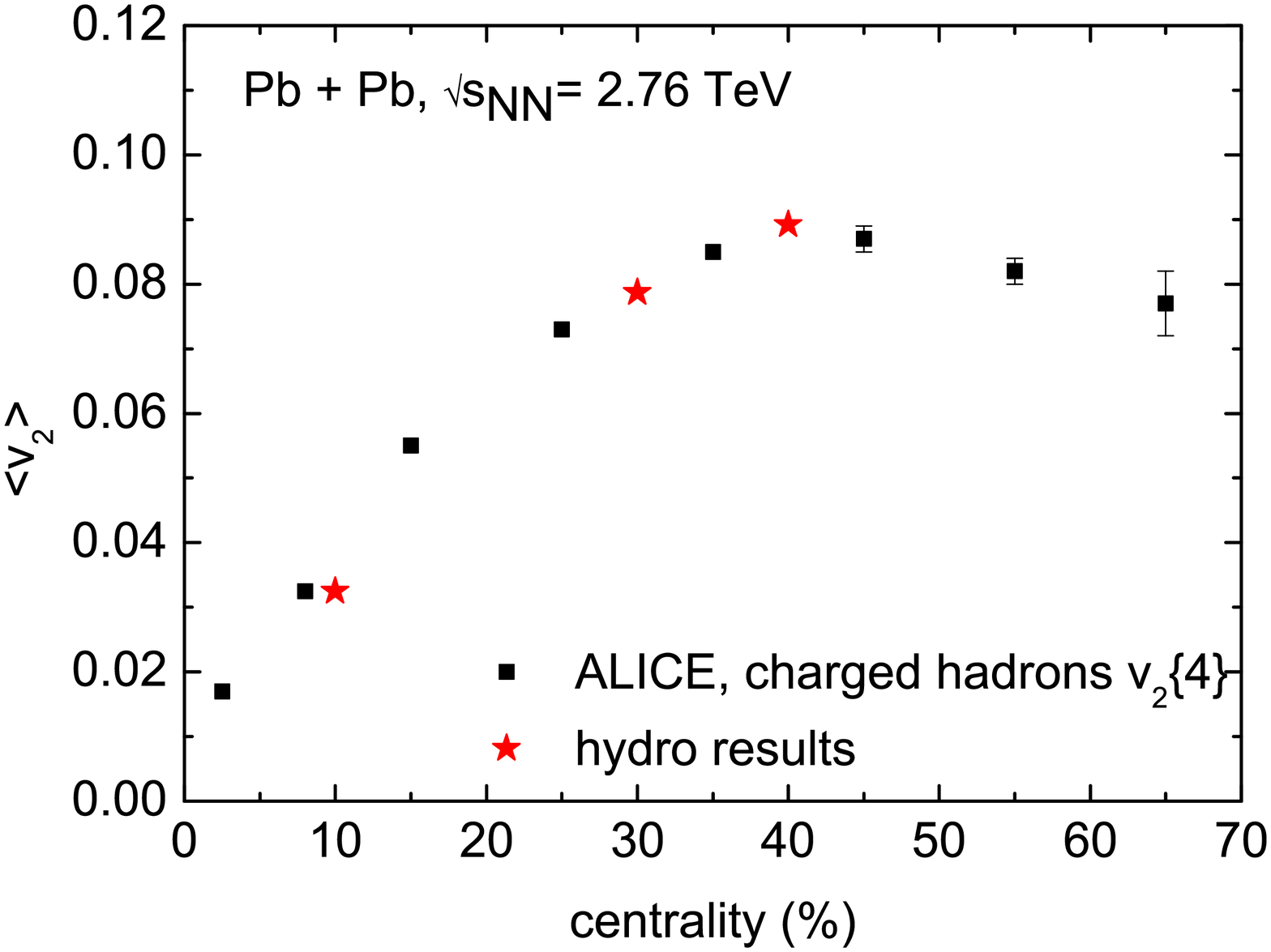}

\vspace{-0.3cm}

\caption{(Color online) Hydrodynamic fits of charged-hadron spectra and
inclusive elliptic flow to ALICE data~\cite{Aamodt:2010jd,Collaboration:2011yba} using our
updated AZHYDRO tune.
}
\label{fig_chargedhadrons}
\end{figure}

For the space-time evolution of the medium in
Pb+Pb($\sqrt{s_{NN}}$=2.76\,TeV), within which HF particles diffuse,
we have retuned the ideal AZHYDRO code~\cite{Kolb:2003dz}. As before, we
employ a lQCD equation of state (EoS)~\cite{Borsanyi:2010cj,Bazavov:2011nk}
with pseudo-critical deconfinement temperature of $T_{\rm pc}=170$\,MeV,
matched to a hadron resonance gas EoS with chemical freezeout
at $T_{\rm ch}=160$\,MeV. Our update pertains to initial conditions for
which we use the Glauber model as in Ref.~\cite{Qiu:2011hf} with an initial
time of 0.4\,fm/$c$, without initial flow nor fluctuations. This yields a
softer expansion than the one adopted
in our previous LHC HF predictions~\cite{He:2012xz}, while the measured
charged-hadron $p_T$ spectra~\cite{Aamodt:2010jd} and inclusive (low-$p_T$)
elliptic flow~\cite{Collaboration:2011yba},
$\langle v_2\rangle$, at kinetic freezeout, $T_{\rm kin}=110$\,MeV, are
fairly well reproduced, cf.~Fig.~\ref{fig_chargedhadrons}. We believe
that the inclusive $v_2$ of charged hadrons, as a measure of the total
bulk momentum anisotropy, provides a suitable calibration for the
calculation of the HF elliptic flow acquired through the coupling to
the medium.

Finally, we have to specify the initial conditions for the HQ spectra. We
replace our previous PYTHIA tune with $\delta$-function fragmentation by a
full FONLL calculation for HQ spectra (using the pertinent software
package~\cite{Cacciari:2001td}) and fragmentation functions (FFs)
according to Ref.~\cite{Braaten:1994bz} for charm (with parameter $r=0.1$),
and Ref.~\cite{Kartvelishvili:1977pi} for bottom (with parameter $\alpha=29.1$).
This framework successfully describes HF spectra in $pp$
at collider energies~\cite{Cacciari:2005rk,Cacciari:2012ny}.
For applications in Pb+Pb we first generate HQ spectra for $pp$ collisions
at $\sqrt{s}$=2.76\,TeV and then fold in the EPS09 shadowing
correction~\cite{ALICE:2012ab,Eskola:2009uj} for charm quarks (but not for bottom).
The resulting spectra are used as the initial condition for the Langevin
simulations of HQ diffusion in the QGP, sampled via the test particle
method. The FONLL fragmentation is also used in the hadronization
process for $c$ and $b$ quarks which do not undergo resonance recombination
at $T_{\rm c}$.

\section{HF Observables at LHC}
\label{sec_obs}
We are now in a position to compute HF observables based on our final $D$-
and $B$-meson spectra in Pb+Pb, i.e., their nuclear modification factor,
\begin{equation}\label{RAA}
R_{\rm AA}(p_T)=\frac{dN_{\rm AA}/dp_Tdy}{N_{\rm coll}dN_{\rm pp}/dp_Tdy} \ ,
\end{equation}
and elliptic flow coefficient,
\begin{equation}\label{v2}
v_2(p_T)=\left\langle \frac{p_x^2 - p_y^2}{p_x^2 + p_y^2} \right\rangle \ ,
\end{equation}
where $N_{\rm coll}$ is the number of binary nucleon-nucleon collisions for
the centrality bin under consideration.

\begin{figure} [!t]
\includegraphics[width=1.05\columnwidth]{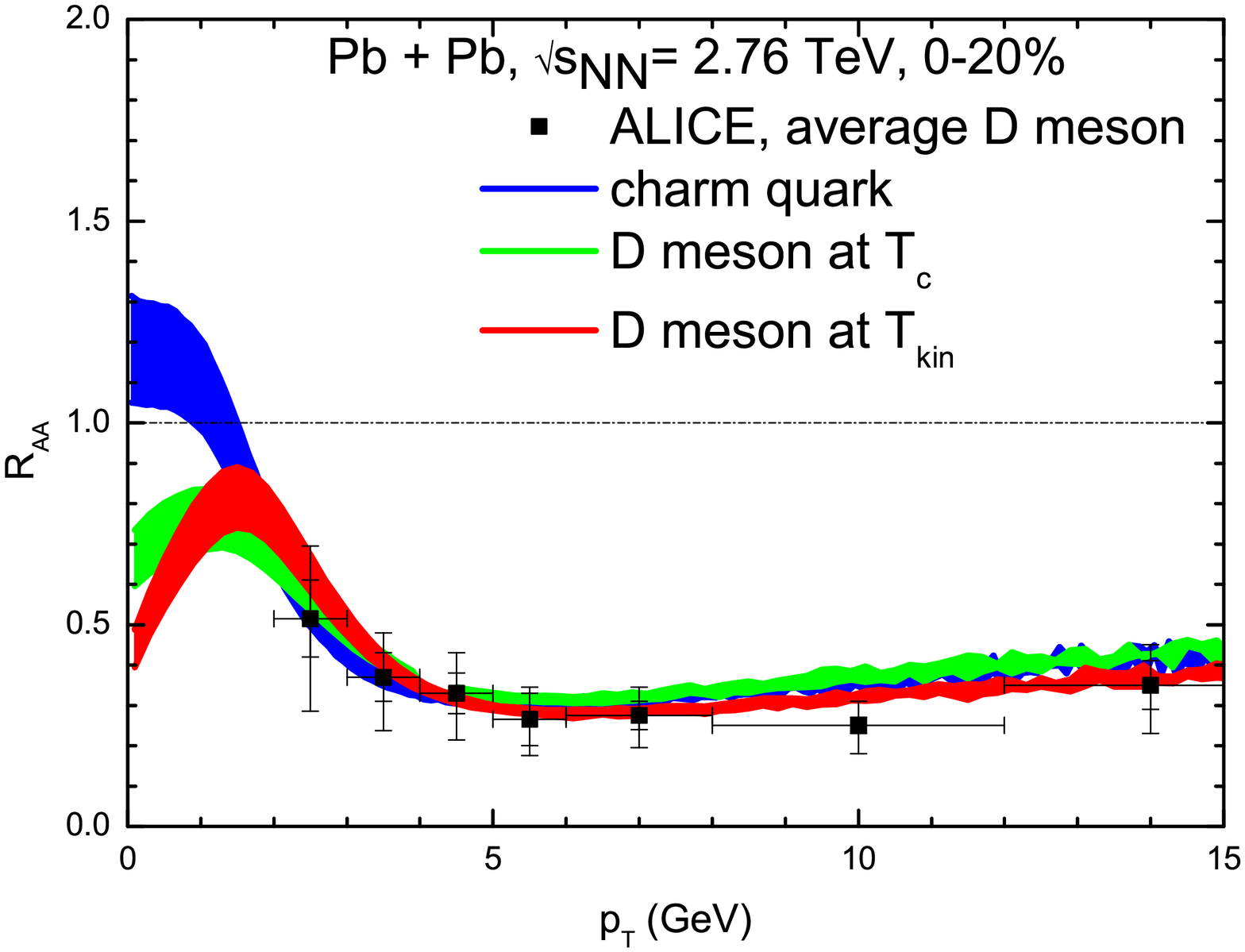}

\vspace{-0.5cm}

\includegraphics[width=1.05\columnwidth]{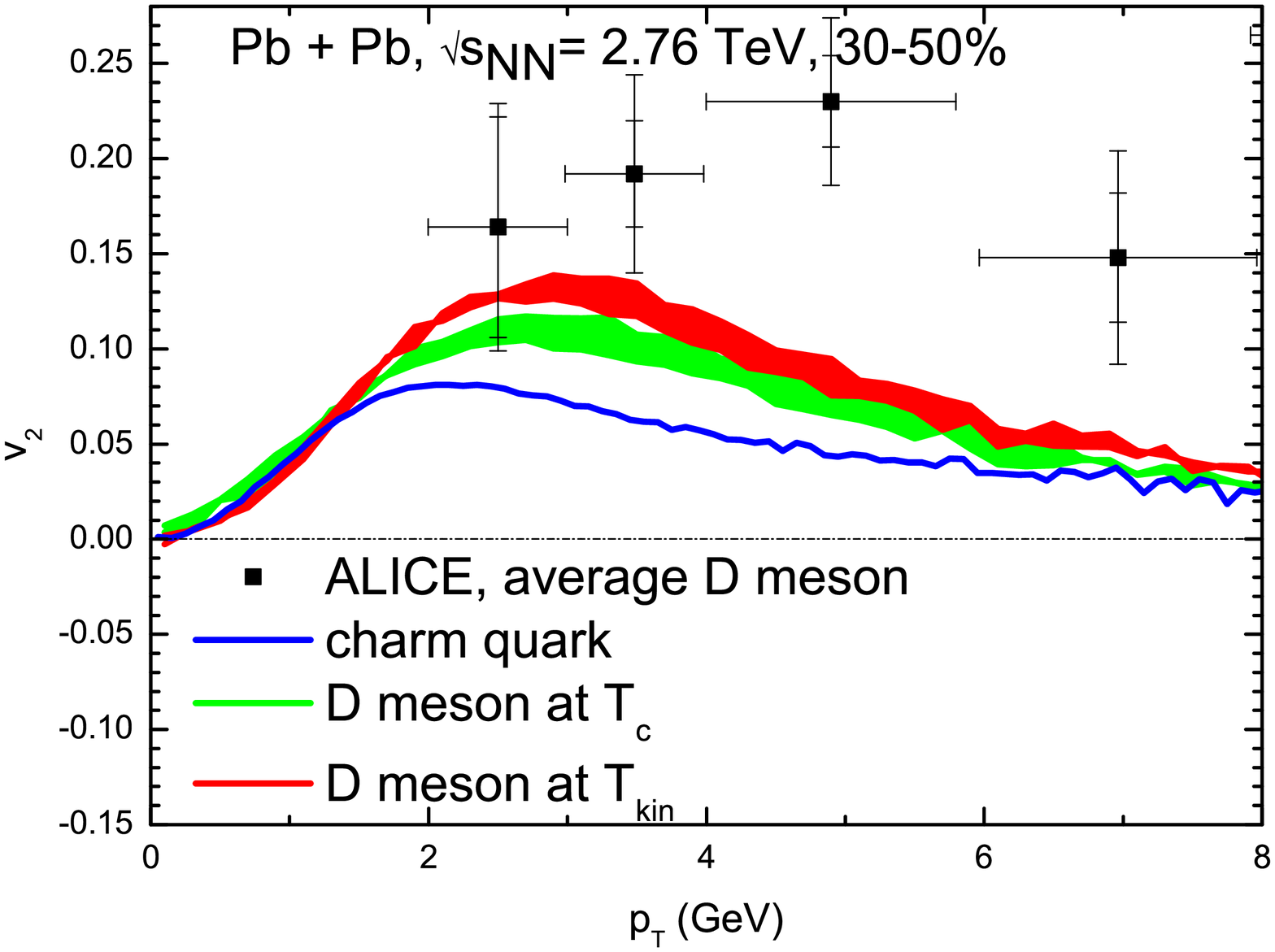}

\vspace{-0.3cm}

\caption{(Color online) $R_{\rm AA}$ (0-20\% Pb+Pb, upper panel) and $v_2$
(30-50\% Pb+Pb, lower panel) for charm quarks and $D$ mesons, compared to
ALICE data~\cite{ALICE:2012ab,Abelev:2013lca}.
For $R_{\rm AA}$ ($v_2$), the bands indicate uncertainties due to shadowing of
charm production (the total charm-quark coalescence probability).}
\label{fig_Dmeson}
\end{figure}
Figure \ref{fig_Dmeson} displays the $R_{\rm AA}$ (0-20\% centrality,
upper panel) and $v_2$ (30-50\%, lower panel) of $c$ quarks (just before
hadronization) and $D$ mesons (just after hadronization and at kinetic
freezeout). Each quantity is shown as a band which encompasses the
leading respective uncertainty, i.e., a shadowing reduction for
$R_{\rm AA}$ (at 64-76\% of the integrated yield), and the recombination
probability for $v_2$ (at 50-85\% for the integrated $c$-quark fraction).
Several features are
noteworthy. The non-perturbative $c$-quark diffusion in the QGP alone
(via the $T$-matrix interaction) brings the $R_{\rm AA}^c$ already
near the ALICE data~\cite{ALICE:2012ab} at intermediate and high $p_T$. Its increasing trend
with $p_T$, resembling the data, is due to the dynamical momentum
dependence of the relaxation rate. At low $p_T$, $R_{\rm AA}^c$ increases
monotonously down to $p_T=0$, indicating the approach to thermalization.
Upon resonance recombination with light quarks around $T_{\rm c}$
the monotonous increase transforms into a flow ``bump" at
$p_T\simeq 1.5$\,GeV in the $D$-meson $R_{\rm AA}$, highlighting the role
of recombination processes as further interactions contributing to
thermalization~\cite{He:2011qa}.
Evidence of a flow bump has been observed at RHIC~\cite{Xie:2013iaa},
corroborating the strong coupling of HF to the medium. At the LHC,
low-$p_T$ ALICE data will thus provide another critical test of the degree
of thermalization in general, and of model predictions to quantify the
magnitude of the HF transport coefficients in particular. Interactions of
$D$ mesons in the hadronic phase have a rather small effect on their final
$R_{AA}$.

\begin{figure} [!t]
\includegraphics[width=1.05\columnwidth]{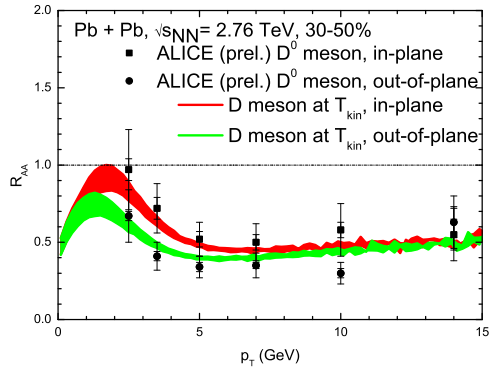}

\vspace{-0.3cm}

\caption{(Color online) $D$-meson in-plane versus out-of-plane $R_{\rm AA}$ for 30-50\% centrality,
compared to ALICE data~\cite{Caffarri:2012wz}. The bands indicate shadowing uncertainties.}
\label{fig_DinvsoutofplaneRAA}
\end{figure}
\begin{figure} [!t]
\includegraphics[width=1.05\columnwidth]{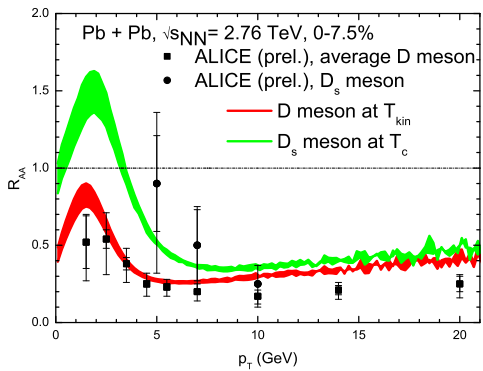}

\vspace{-0.3cm}

\caption{(Color online) $D$- versus $D_s$-meson $R_{\rm AA}$ for 0-7.5\%
central Pb+Pb, compared to ALICE data~\cite{Innocenti:2012ds}.
The bands indicate the charm-shadowing uncertainty.}
\label{fig_DvsDsRAA}
\end{figure}

The situation is somewhat different for the elliptic flow. The combined
effect of hadronization and hadronic diffusion increases the peak value of
the final $D$-meson $v_2$ by $\sim$75\% over the QGP induced $c$-quark $v_2$.
On the one hand, this is due to the bulk-$v_2$ taking a few fm/$c$ to build
up, and, on the other hand, due to the small spatial diffusion coefficient,
${\cal D}_s$$\simeq$~3-4/$(2\pi T)$, around $T_{\rm pc}$ (on both QGP and
hadronic side). The prominent role of resonance recombination as an
interaction driving HF toward equilibrium is once
again apparent. This increase in $v_2$, which at the same time affects
the $R_{AA}$ relatively little, appears to be an important ingredient
to simultaneously describe the ALICE data for {\em both} observables.
Our calculation comes close to the combined $D$-meson $v_2$ data from
ALICE~\cite{Abelev:2013lca} up to $p_T\simeq4$\,GeV, while it falls below above.
This is reiterated by comparing our results
to the in- vs. out-of-plane $R_{AA}$ data~\cite{Caffarri:2012wz}: the splitting tends to be
underestimated at high $p_T$, cf.~Fig.~\ref{fig_DinvsoutofplaneRAA}. In fact, the suppression observed in the
$R_{AA}$ at high $p_T$ for the most central data sample (0-7.5\%) is
also significantly underestimated by our calculations,
cf.~Fig.~\ref{fig_DvsDsRAA}.
All of this points toward a lack of path-length dependence in our elastic
quenching mechanism at high $p_T$, for which radiative energy loss is a
natural candidate. Its rigorous implementation into a transport framework,
including interference effects, currently remains a challenge.


The $D_s$-meson $R_{\rm AA}$ and $v_2$ at low and intermediate $p_T$ have
recently been proposed as a remarkable signature to quantitatively probe the role
of $c$-quark recombination and hadronic diffusion in URHICs~\cite{He:2012df}.
An enhancement of the $D_s$ over the $D$ $R_{AA}$ has been
predicted as a consequence of the well-established strangeness enhancement
in URHICs (relative to $pp$ collisions), realized through $c$-quark
recombination with equilibrated strange quarks in the
QGP~\cite{Kuznetsova:2006bh}. Our predictions for LHC are compared to
preliminary ALICE data~\cite{Innocenti:2012ds} in Fig.~\ref{fig_DvsDsRAA},
which indeed give a first indication of the proposed enhancement.
At high $p_T$, fragmentation (universal in $pp$ and Pb+Pb) leads to similar
$R_{\rm AA}$'s for $D$ and $D_s$ mesons, with a small splitting induced
by an extra suppression of $D$ mesons due to interactions in the hadronic
phase; for $D_s$ mesons hadronic rescattering is believed to be
small and has been neglected in our calculations~\cite{He:2012df}.

\begin{figure} [!t]
\includegraphics[width=1.05\columnwidth]{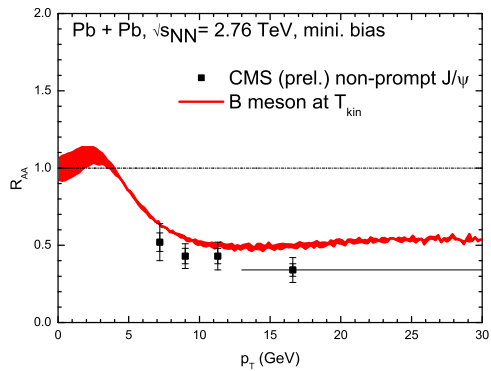}

\vspace{-0.3cm}

\includegraphics[width=1.05\columnwidth]{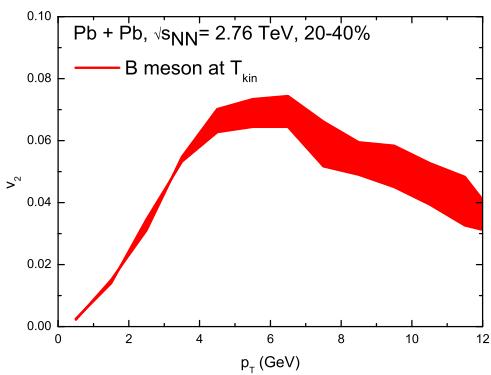}
\caption{(Color online) $B$-meson $R_{\rm AA}$ (upper panel) and $v_2$
(lower panel) in minimum-bias Pb+Pb. The bands indicate the uncertainty in the
total $b$-quark coalescence probability (no shadowing is
applied). The CMS data in the upper panel~\cite{Chatrchyan:2012np,CMS:2012wba} are
for non-prompt $J/\psi$ (associated with $B$ decays) plotted vs.~the $J/\psi$ $p_T$
(no rescaling for $B\to J/\psi+X$ decays is applied).}
\label{fig_Bmeson}
\end{figure}
Next we turn to the bottom sector. Current information on $B$-meson
spectra in Pb+Pb collisions is available through the CMS measurements of
non-prompt $J/\psi$'s associated with $B\rightarrow J/\psi + X$
decays~\cite{Chatrchyan:2012np,CMS:2012wba}. We calculate the $B$-meson $R_{\rm AA}$
for minimum bias Pb+Pb from a $N_{\rm coll}$-weighted average over the three
centrality bins 0-10\%, 20-40\% and 50-80\%, see upper panel of Fig.~\ref{fig_Bmeson}.
Since we do not introduce any shadowing correction for bottom, the
uncertainty band in both $R_{\rm AA}$ and $v_2$ refers to the integrated
recombination probability of $\sim$50-90\%. At low $p_T$, the $B$-meson
$R_{\rm AA}$ is close to 1 with a small flow ``bump", i.e., a maximum at
finite $p_T\simeq$~2-3\,GeV, while the suppression for $p_T\gsim10$\,GeV
is rather flat at $\sim$0.5. This is roughly consistent with the CMS
non-prompt $J/\psi$ data (we made no attempt to rescale the
$J/\psi$ momenta to reflect the parent $B$-meson momenta). The $B$ mesons
also acquire a sizeable $v_2$, reaching up to 7.5\%, implying
a significant approach to thermalization of bottom through diffusion and
resonance recombination with light quarks. In contrast to charm, the
bottom $v_2$ peaks at a much higher $p_T$, which is in part a kinetic mass
effect, but also due to a flatter momentum dependence of the $b$-quark
relaxation rate~\cite{Riek:2010fk,Huggins:2012dj} and a coalescence
probability function $P_{\rm coal}(p_t^Q)$ decreasing more
slowly than for charm~\cite{He:2011qa}.

\begin{figure} [!t]
\includegraphics[width=1.05\columnwidth]{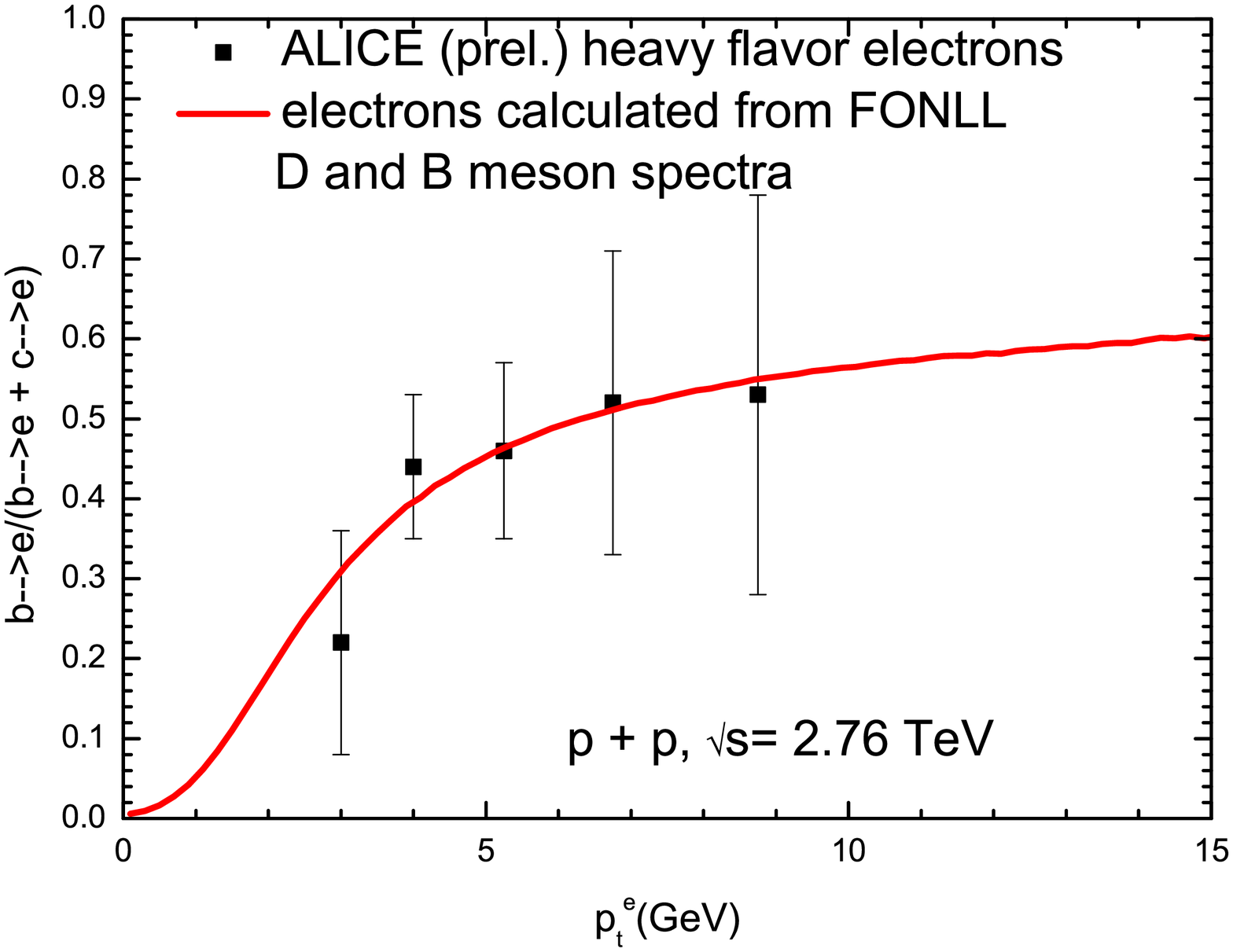}

\vspace{-0.3cm}

\caption{(Color online) Ratio of electrons from bottom to charm+bottom
in $pp$ collisions in comparison with ALICE data~\cite{Kweon:2012yn},
assuming a total bottom-to-charm cross section ratio of $\sim 0.05$.}
\label{fig_cvsbelectron}
\end{figure}

\begin{figure} [!t]
\includegraphics[width=1.05\columnwidth]{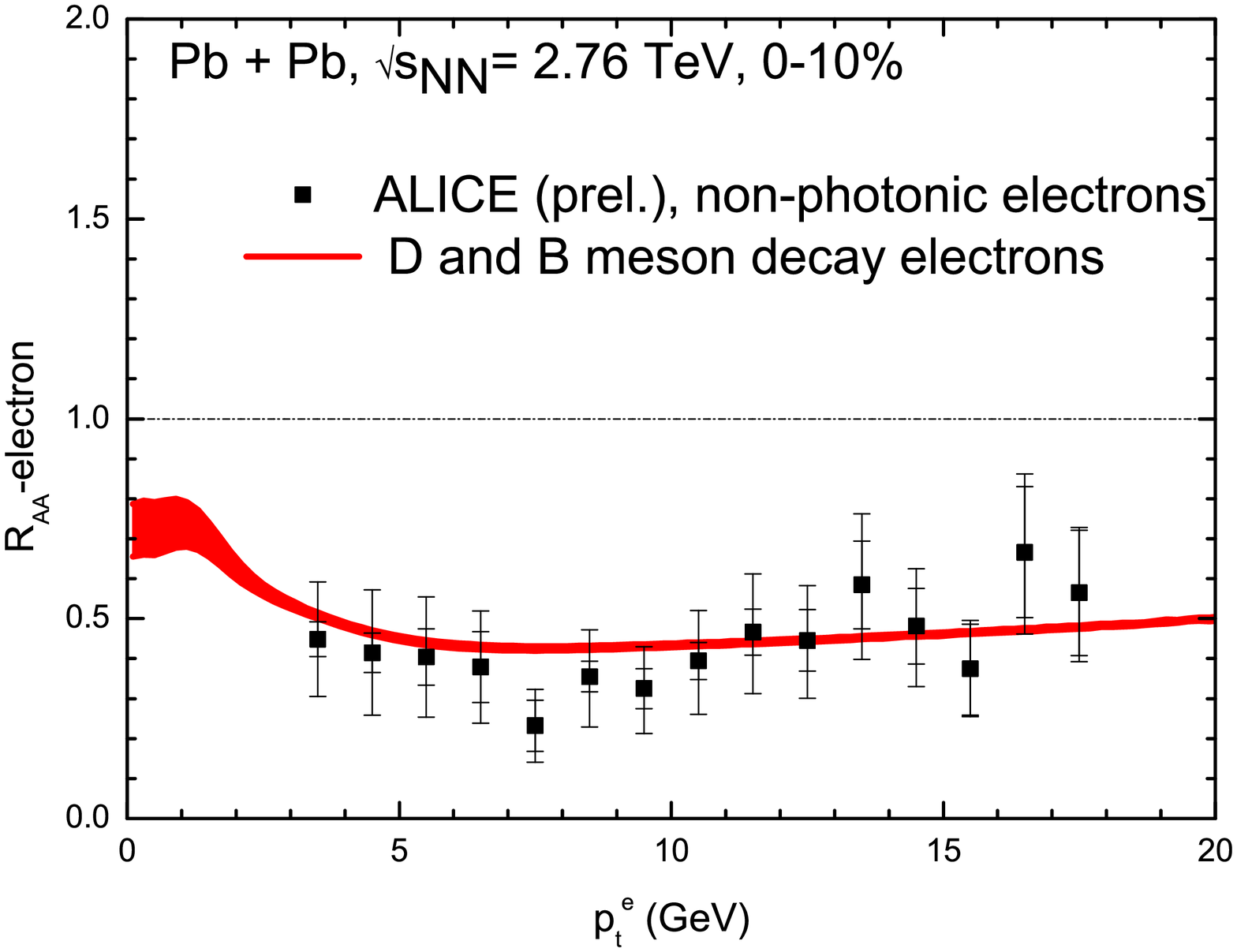}

\vspace{-0.5cm}

\includegraphics[width=1.05\columnwidth]{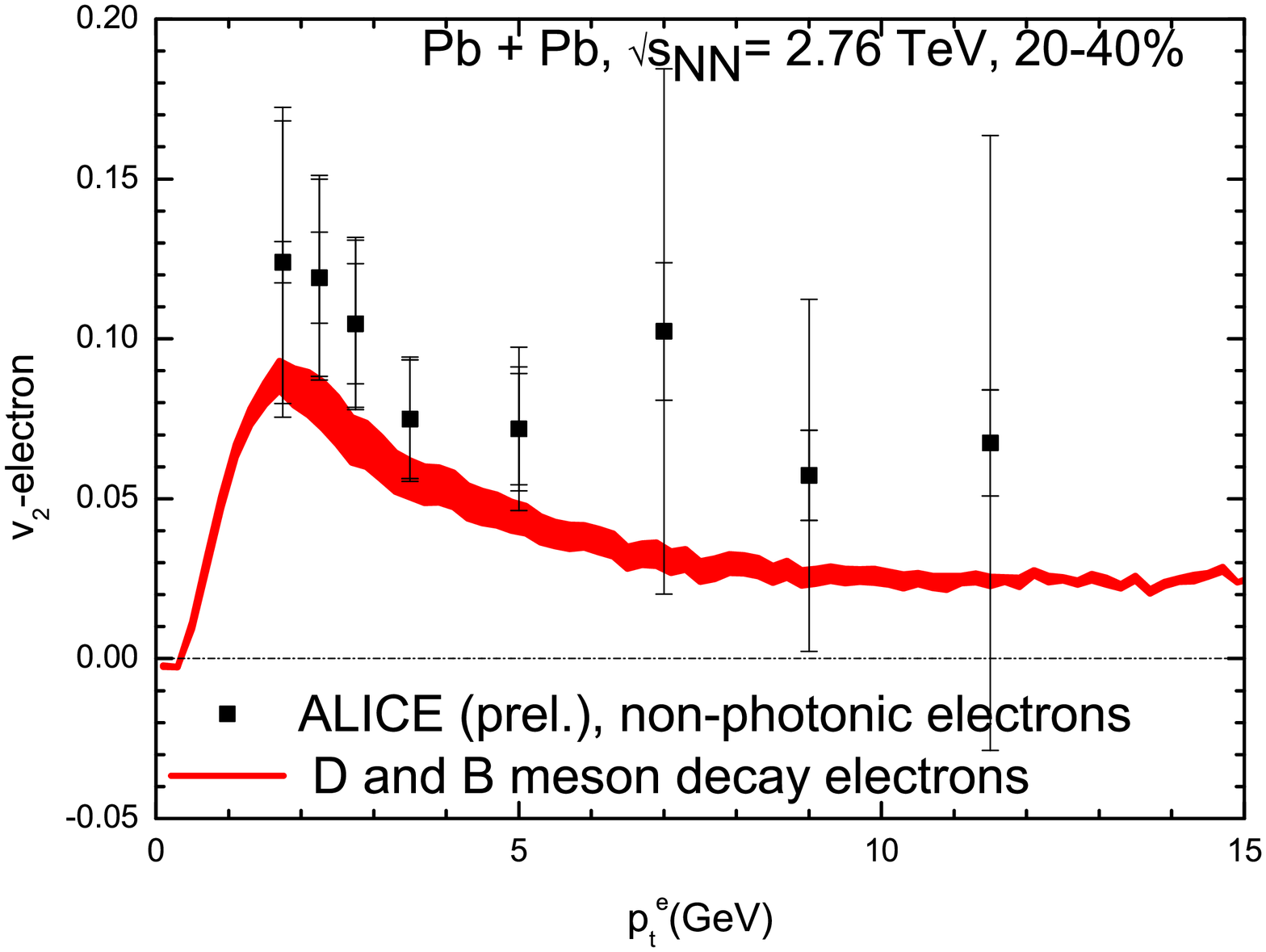}

\vspace{-0.3cm}

\caption{(Color online) Heavy-flavor electron $R_{\rm AA}$ and $v_2$, compared
to ALICE data~\cite{Sakai:2013ata}. For $R_{\rm AA}$, $\sim$50\% coalescence probability
is applied for both $D$ and $B$ and the band indicates uncertainty in charm shadowing.
For $v_2$ the band indicates the uncertainty due to $c$- and $b$-quark
coalescence probabilities of $\sim$50-90\%.}
\label{fig_HFelectros}
\end{figure}
Finally, we compute HF electron observables from the semi-leptonic decays
of $D$ and $B$ mesons. We first determine the bottom fraction of the
single electrons in the $pp$ baseline, which is illustrated in
Fig.~\ref{fig_cvsbelectron}. With a bottom-to-charm cross section ratio
of $\sim$0.05, the pertinent ALICE data~\cite{Kweon:2012yn}
can be reproduced. In our calculation
the bottom contribution exceeds the charm one for electron momenta above
$p_t^e\simeq6$\,GeV,
although the ratio becomes quite flat. With this input we can convert our
$D$- and $B$-meson observables computed above into single-electron ones,
cf.~Fig.~\ref{fig_HFelectros} (as before, the bands indicate the leading
uncertainties, i.e., charm shadowing for $R_{\rm AA}$ and charm/bottom
integrated coalescence probabilities for $v_2$). The ALICE data~\cite{Sakai:2013ata} for
$R_{\rm AA}^e$ are reasonably well described while the calculated
electron-$v_2$ appears to be somewhat low, especially toward higher
$p_t^e$. In this regime the $v_2^e$ is largely determined by the $B$-meson
$v_2$ as shown in the lower panel of Fig.~\ref{fig_Bmeson}. These features
confirm the trends of the individual $D$- and $B$-meson observables.

\section{Summary}
\label{sec_sum}

We have presented a comprehensive study of open HF probes in Pb+Pb
collisions at $\sqrt{s_{\rm NN}}=2.76~{\rm TeV}$ using a nonperturbative
transport model which implements a strong-coupling approach of heavy quarks
and hadrons into a hydrodynamically expanding medium. An overall fair
description of the current data set from ALICE and CMS on the nuclear
modification factor and elliptic flow of $D$, $D_s$, non-prompt $J/\psi$
from $B$ decays and HF leptons emerges. In particular, our approach eases the
tension between $R_{AA}$ and $v_2$ found previously, helped by a modest
update of our QGP transport coefficient (now including nonperturbative
HQ-gluon interactions) and a ``softer" hydro expansion due to
modified initial conditions. The key mechanism, however, first found in
Ref.~\cite{vanHees:2005wb}, is a strong HF coupling to the medium around
$T_{\rm pc}$ (on {\em both QGP} and hadronic side), which includes the
effects of resonance recombination. This insight corroborates that the QCD
medium is most strongly coupled in the quark-to-hadron transition region
of the phase diagram.
At higher $p_T$ our purely elastic treatment of HF-medium interactions seems
to lack some strength and path-length dependence, which is not unexpected.
On the other hand, more precise data at low and intermediate $p_T$, where
we believe our approach to be most reliable, will allow for quantitative
tests of the HF transport properties and their origin. Future plans include
the study of HF baryons and more differential observables like
HF correlations~\cite{Zhu:2006er,Zhu:2007ne,Nahrgang:2013saa}.
The inclusion of radiative effects will be required to improve the
phenomenology at high momenta.\\

{\bf Acknowledgments:}
We are indebted to F.~Riek and K. Huggins for providing
the results for the HQ transport coefficients. This work was supported
by the U.S.~National Science Foundation (NSF) through CAREER grant PHY-0847538
and grant PHY-1306359, by the A.-v.-Humboldt
Foundation, by the JET Collaboration and DOE grant DE-FG02-10ER41682,
and by NSFC grant 11305089.

\end{document}